\documentclass[twocolumn,showpacs,preprintnumbers,amsmath,amssymb]{revtex4}

\usepackage{graphicx}
\usepackage{amsmath,amsfonts}
\begin{document}

\author{D. Kleinhans} \author{R. Friedrich}
\affiliation{Westf{\"a}lische Wilhelms-Universit{\"a}t M{\"u}nster,
  Institut f\"ur Theoretische Physik, D-48149 M{\"u}nster, Germany}

\title{Continuous Time Random Walks: Simulation of continuous
  trajectories} \date{\today}

 

\begin{abstract}
  Continuous time random walks have been developed as a
  straightforward generalisation of classical random walk processes.
  Some 10 years ago, Fogedby introduced a continuous representation of
  these processes by means of a set of Langevin equations [H.~C.
  Fogedby, Phys.~Rev.~E {\bf 50} (1994)].  The present work is devoted
  to a detailed discussion of Fogedby's model and presents its
  application for the robust numerical generation of sample paths of
  continuous time random walk processes.
\end{abstract}

\maketitle

\section{\label{sect:introduction}Introduction}
The analysis of stochastic processes by Bachelier, Einstein and
Langevin \cite{Bachelier00,Einstein05:Brown,Lemons} extensively has
inspired scientist in the last century. First, Fokker-Planck equations
describing the time evolution of probability density functions (pdfs)
emerged. In addition, mathematical methods for proper interpretation
of the associated Langevin equations have been developed, that allow
access to the trajectories of individual particles. An intrinsic
feature of these processes is, that they obey Markov properties
\cite{Risken}. Therefore, two point statistics are sufficient for a
complete description of these processes.

In recent years, processes exhibiting anomalous diffusion,
$\left\langle x^2(t)\right\rangle\sim t^\xi$ with $\xi\ne 1$,
increasingly have attracted attention \cite{Metzler00}. Such processes
typically are realised in complex environments such as porous and
disordered media, see e.g.~\cite{Dentz04} and references therein. In
contrast to ordinary diffusion, Markov properties do not hold for
these processes. Therefore, multipoint joint statistics have to be
considered for proper description of the dynamics.  Likewise, two
alternative approaches to these processes have been evolved. On the
one hand, these processes can be described by means of fractional
Fokker-Planck equations, that contain fractional derivatives with
non-local character. On the other hand, Continuous Time Random Walk
(CTRW) processes \cite{Montroll65} have been proposed for the analysis
of the microscopic properties of anomalous diffusion processes
\cite{Klafter87}.  In general, they are specified by the iterative
discrete equations \cite{Weiss,Metzler00}
\begin{subequations}
  \label{eqn:ctrw-disc}
  \begin{eqnarray}
    x_{i+1}&=&x_i+\eta_i\\
    t_{i+1}&=&t_i+\tau_i\quad,
  \end{eqnarray}
\end{subequations}
where $(\eta_i,\tau_i)$ is a set of random numbers drawn from the pdf
$\Psi(\eta,\tau)$, that vanishes for negative values of $\tau$ for
reasons of causality.  Frequently, equations (\ref{eqn:ctrw-disc}) are
used to model time-continuous processes with the additional assignment
\cite{Montroll65, Weiss}
\begin{equation}
  x(t)=x_i \quad\mbox{with}\quad t_i\le t<t_{i+1}\quad.
\end{equation}
With the aid of CTRWs, limiting behaviour and ensemble statistics of
anomalous diffusion processes become accessible through Monte-Carlo
simulations. For details, the reader is referred to recent works by
Dentz et al.~\cite{Dentz04}, Heinsalu et al.~\cite{Heinsalu06} and
Gorenflo et al.~\cite{Gorenflo07}.

Recently, Fogedby formulated a \emph{continuous} description of CTRWs,
that is based of a set of stochastic differential equations
\cite{Fogedby94},
\begin{subequations}
  \label{eqn:fogedby}
  \begin{eqnarray}
    \label{eqn:fogedby-x}    \frac{dx}{ds}&=&F(x)+\eta(s)\\
    \label{eqn:fogedby-t}    \frac{dt}{ds}&=&\tau(s)\quad.
  \end{eqnarray}
\end{subequations}
Here, the discrete variable $i$ of equations (\ref{eqn:ctrw-disc}) is
generalised to the continuous variable $s$, that can be associated
with an intrinsic time of the CTRW.  A number of publications
addressed statistical properties of trajectories of this approach
\cite{Baule05,Baule07EPL,Baule07}.  It is the aim of the present work,
to present a robust algorithm for the generation of \emph{continuous}
sample paths from Fogedby's equations.  By this means trajectories can
be obtained, that properly exhibit the anomalous dynamics of CTRWs on
any time scale.

This work is structured as follows. In the next section, some general
remarks are made on the definition of continuous CTRWs, equations
(\ref{eqn:fogedby}). Section \ref{sect:levy-prop} is dedicated to the
properties of the process $t(s)$, equation (\ref{eqn:fogedby-t}), that
typically is driven by Levy noise. The numerical simulation of
continuous CTRW processes is presented in section \ref{sect:ctrw-sim}
and exemplified by means of some results in section \ref{sect:exa}. We
conclude with section \ref{sect:conclu}, that summarizes our results
and suggests future applications.

\section{\label{sect:fogedby}CTRWs in the spirit of Fogedby: Some
  remarks}
First of all, the character of the distributions of the random
variables $\eta$ for the jump length and $\tau$ for the waiting time
has to be addressed. In case of the discrete definition, equations
(\ref{eqn:ctrw-disc}), a broad class of distributions is feasible for
this purpose. The continuous Langevin formulation of Fogedby,
equations (\ref{eqn:fogedby}), however, requires the associated
distributions to be stable in order to be properly defined.  For a
short introduction into the concept of stable distributions we refer
to \cite{Metzler00}. In the long time limit, however, both approaches
are equivalent.

In 1994, Fogedby introduced the stochastic differential equations
(\ref{eqn:fogedby}) for CTRWs as \emph{the continuum limit of the path
  parameter or arc length $s$ along the trajectory} \cite{Fogedby94}.
He considered independent random variables $\eta(s)$ and $\tau(s)$
with power law behaviour, that is
\begin{subequations}
  \begin{eqnarray}
    \Psi(\eta,\tau)\sim
    \eta^{-1-\xi_\eta}\tau^{-1-\xi_{\tau}}\quad\mbox{for}\quad
    \eta,\tau\gg 1\quad.
  \end{eqnarray}
\end{subequations}
For reasons of normability, Fogedby used cutoffs at low values of
$\left|\eta\right|$ and $ \tau$, respectively. He mainly was
interested in the properties of the process $x(t)$, that can be
obtained from equations (\ref{eqn:fogedby}) by inversion of the latter
process. By means of his modified power-law distributions, the
long-time behaviour of processes with power law waiting jump length
and waiting time distributions could be derived. In this context,
first properties of the inverse process $s(t)$ of equation
(\ref{eqn:fogedby-t}) have been addressed.

Baule et al.~investigated the properties of the inverse process $s(t)$
in greater detail \cite{Baule05}. In particular, multi-time joint
probabilities could be calculated. The waiting times $\tau$ were
considered to obey one-sided Levy distributions with tail parameter
$\alpha$, that are discussed in greater detail in the following
section.  In absence of external force terms ($F=0$ in equation
(\ref{eqn:fogedby-x})) analytical expressions for correlations
functions could be derived by application of the inverse Fourier- and
Laplace-transforms. Recently, these results could be extended to
Ornstein-Uhlenbeck like processes with a linear repelling term, thus
$F(x)=-\gamma x$ \cite{Baule07}.

Both works, however, focus on the ensemble statistics, whereas the
Langevin approach of Fogedby, that is interesting itself, has not
attracted much attention yet.

\section{\label{sect:levy-prop} Properties of $t(s)$ and its inverse
  process $s(t)$}
In this section, we concentrate on the process $t(s)$.  Due to
causality, this process has to be strictly monotonically increasing.
The increments $dt(s)=t(s+ds)-t(s)$, therefore, have to obey fully
skewed stable distributions.

\begin{figure}
  \includegraphics*[width=8cm]{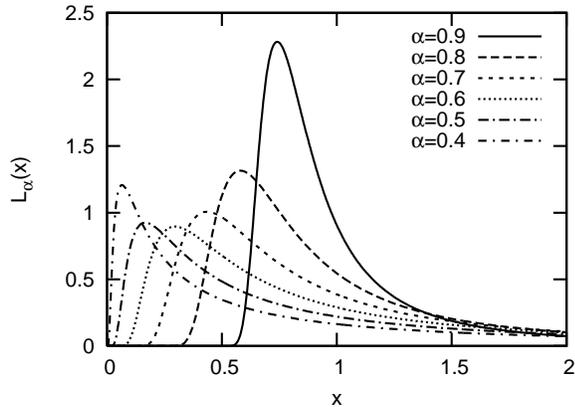}
  \caption{\label{fig:levybeisp}Examples for the fully skewed Levy
    distribution according to equation (\ref{eqn:mylevy}) for
    different characteristic exponents $\alpha$. For $\alpha\to 1$,
    the pdf converges to $\delta(x-1)$.}
\end{figure}

Generally, stable distribution are assigned to the $\alpha$-stable
probability distribution functions (pdfs). $\alpha$-stable Levy
distributions typically are not available in a closed form but are
only accessible by means of their Fourier transform. We especially
consider the distribution
\begin{equation}
  \label{eqn:mylevy}
  L_{\alpha}(x)=\frac{1}{\pi}Re\left\{\int\limits_{0}^{\infty}dz\
    \exp\left(-ikx-z^\alpha\exp\left[-i\frac{\alpha\pi}{2}\right]\right)\right\}\quad.
\end{equation}
Here, $i$ is the imaginary unit and $0<\alpha\le 1$ the stability
index, that specifies the asymptotic behaviour $L_{\alpha}(x)\sim
x^{-(1+\alpha)}$ at $x\gg 1$. This pdf complies with a common
parametrization of $\alpha$-stable pdfs
\cite{Metzler00,Janicki,Weron01},
\begin{eqnarray}
  L_\alpha^{\beta,c}(x)&=&\frac{1}{2\pi}\int\limits_{-\infty}^{\infty}dz\ \\\nonumber&&
  \exp\left(-ikx-c z^\alpha\left[1-i\beta\frac{z}{|z|}\tan\frac{\alpha\pi}{2}\right]\right),
\end{eqnarray}
for the parameters $\beta=1$ and
$c=\left(1+\tan^2\frac{\alpha\pi}{2}\right)^{(-1/2)}$.  Some examples
for this distribution are depicted in figure \ref{fig:levybeisp}. An
important feature of this representation is, that it according to
equation (\ref{eqn:mylevy}) is defined even for $\alpha=1$ by means of
$L_{\alpha=1}(x)=\delta(x-1)$. From the theory of stable processes it
follows, that the increment $dt$ has to obey the distribution
$ds^{-\alpha}L_\alpha(dt/ds^{\alpha})$ \cite{Janicki}.  The case
$\alpha=1$ with $dt=ds$ complies with a stochastic processes, that
solely can be described by equation (\ref{eqn:fogedby-x}) with $s=t$.

\begin{figure}
  \includegraphics*[width=8cm]{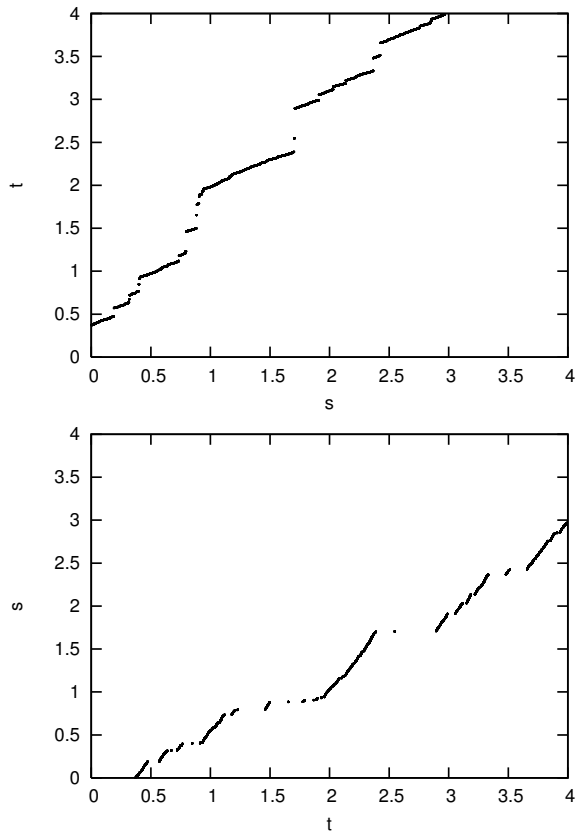}
  \caption{\label{fig:plot-st-ts}Trajectories of the process $t(s)$
    (upper panel) and the associated inverse process $s(t)$ (lower
    panel) simulated for $\alpha=0.9$. The finite jumps of the process
    $t(s)$, that are characteristic for Levy processes, are evident.
    Due to these jumps the inverse process $s(t)$, however, a priori
    is not defined on $\left[0,\infty\right[$.}
\end{figure}

An intrinsic feature of Levy processes is, that trajectories contain
finite jumps in terms of discontinuities with a probability greater
than zero. They are continuous only from the right
\cite{Meerschaert04}, that is
\begin{equation}
  \lim\limits_{\Delta s\to 0} t(s+\Delta s)=t(s)\quad .
\end{equation}
A sample of a fully skewed Levy process with characteristic exponent
$\alpha=0.9$ is depicted in the upper panel of figure
\ref{fig:plot-st-ts}. Due to these jumps, the range of values $t(s)$
does not cover the full interval $[0,\infty[$. Rather, the range can
be specified by means of a set of intervals. Due to the monotonic
increase of the process $t(s)$ a inverse process $s(t)$ exists. This
process has been applied for construction of the process
$x(t)=x(s(t))$ in the past. It, however, a priori is not properly
defined for $t\in [0,\infty[$ due to the jump characteristic of the
Levy process. A sample of the inverse process is depicted in the lower
panel of figure \ref{fig:plot-st-ts}.

In general, the meaning of the inverse function has to be specified in
order to be properly defined on $t\in [0,\infty[$.  Appropriate
definitions for the inverse function e.g.~are \cite{Meerschaert04}
\begin{subequations}
  \label{eqn:invers-def}
  \begin{eqnarray}
    \label{eqn:invers-def-inf}s(t)&:=&\inf\left\{s:t(s)\ge t\right\}\quad\mbox{or}\\
    s(t)&:=&\sup\left\{s:t(s)\le t\right\}\quad.
  \end{eqnarray}
\end{subequations}

If the process $x(s)$, is steady, these definitions are equivalent in
the limit $\Delta s \to 0$ for at $t\in[0,\infty[$: Since $x(s)$ is
steady,
\begin{equation}
  \label{eqn:steady}
  \lim\limits_{\Delta s\to 0}x(s-\Delta s)=\lim\limits_{\Delta s\to 0}x(s+\Delta s)=x(s)
\end{equation}
is valid by definition. Due to the monotonic character of the process
$t(s)$, $t(s)\le t(s+\Delta s)$ for $\Delta s \ge 0$. Consequently,
$s\le s(t')\le s+\Delta s$ if $t(s)\le t'\le t(s+\Delta s)$. Then,
\begin{equation}
  \lim\limits_{\Delta
    s\to 0}x(s)=x(s(t'))=x(s+\Delta s)
\end{equation}
is valid for $t(s)\le t'\le t(s+\Delta s)$. Here, for the process
$t(s)$ only the continuity from the rights has been used.  For steady
processes $x(s)$ in the limit of infinitesimal $\Delta s$, thus, no
additional definition for proper interpretation of the inverse
function is required.

If unsteady jumps of $x(s)$ coincide with those of $t(s)$, the latter
argumentation is not feasible. We, however, restrict to stochastic
processes $x(s)$ with Gaussian noise, that are steady with probability
$1$ for $s\in[0,\infty[$.

\section{\label{sect:ctrw-sim} Numerical simulation of sample paths}

An equivalent formulation of equations (\ref{eqn:fogedby}) is given by
the integral equations
\begin{subequations}
  \label{eqn:int}
  \begin{eqnarray}
    \label{eqn:int-x}    x(s)&=&x(0)+\int\limits_{0}^{s}ds' F(x(s'))+\int\limits_{0}^{s}dW(s')\\
    \label{eqn:int-t}    t(s)&=&t(0)+\int\limits_{0}^{s} dL_{\alpha}(s')\quad,
  \end{eqnarray}
\end{subequations}
where $dW$ and $dL_\alpha$ are the infinitesimal increments of Wiener
and $\alpha$-stable Levy processes, respectively. For numerical
integration these equations have to be discretisized with an adequate
discrete increment $\Delta s$.  Application of the Euler scheme for
numerical evaluation of equations (\ref{eqn:int}) then yields
\begin{subequations}
  \label{eqn:ctrw-fogedby-disc}
  \begin{eqnarray}
    x(s+\Delta s)&=&x(s)+\Delta sF(x(s))+\eta(s,\Delta s)\\
    \label{eqn:ctrw-fogedby-disc-t}
    t(s+\Delta s)&=&t(s)+\tau_{\alpha}(s, \Delta s)\quad.
  \end{eqnarray}
\end{subequations}
Here, the random variables $\eta(s_i,\Delta s)$ independently have to
be drawn from a Gaussian pdf with variance $\sigma^2=\Delta s$. The
variables $\tau_\alpha(s_i, \Delta s)$ have to comply with the
distribution $\frac{1}{\Delta
  s^{\alpha}}L_\alpha(\frac{\tau_\alpha}{\Delta s^{\alpha}})$.  The
efficient numerical generation of these random numbers is addressed in
appendix \ref{sect:randgen}.  Due to the absence of forcing and the
purely additive character of the noise, the integration of $t(s)$ by
means of the Euler scheme is exact.  For numerical integration of the
process $x(s)$ with Gaussian noise, alternatively advanced
discretization schemes can be applied \cite{Kloeden}.

For numerical simulation of trajectories $x(t)$ at discrete times
$t_j:=j\Delta t$, $j=0,\ldots,N$, the inverse $s(t)$ does not have to
be calculated explicitly. Instead, the following algorithm can be
applied, that incorporates definition (\ref{eqn:invers-def-inf}) for
the inverse process:
\begin{itemize}
\item Initialisation of $x_s(0)$ and $t_s(0)$, set $s=0$
\item for every $j=0$ to $N$:
  \begin{enumerate}
  \item while $(t_s(s)< t_j)$:
    \begin{enumerate}
    \item calculate $x_s(s+\Delta s)$ and $t_s(s+\Delta s)$ from
      eqns.~(\ref{eqn:ctrw-fogedby-disc})
    \item increase $s$ by $\Delta s$
    \end{enumerate}
  \item set $x(t_j):=x_s(s)$
  \end{enumerate}
\end{itemize}

The discretization of $t$, $\Delta t$, is given by the desired
sampling rate of the simulated process. The optimal value for the
discretization of the intrinsic variable $s$, $\Delta s$, depends on
characteristic length scales of the process $x(s)$ and the desired
accuracy of the resulting process $x(t)$. Typically, $\Delta s$ has to
be adjusted, such that the right hand side term of the discrete
equation, $\Delta sF(x(s_i))+\eta(s_i,\Delta s)$, with a sufficient
probability is less than the desired accuracy of the simulation. This
argument is illustrated within the scope of the following section. Too
small values for $\Delta s$, in turn, may bias the accuracy of the
numerical evaluation of equations (\ref{eqn:ctrw-fogedby-disc}) due to
discretisation errors and reduce the speed of the algorithms.  The
choice for $\Delta s$ therefore is a trade-off between accuracy of the
discretization, validity of the inversion of the process $t(s)$,
computer time and discretization errors. For details concerning the
numerical evaluation of the discretized equations
(\ref{eqn:ctrw-fogedby-disc}), the reader is referred to the book of
Kloeden and Platen \cite{Kloeden}.

\section{\label{sect:exa}Examples}
For exemplification of the simulation procedure and characteristic
properties of continuous CTRWs, processes with $F(x)=-x$ are
considered. The process $x(s)$ then is an ordinary Ornstein-Uhlenbeck
process
\begin{equation}
  dx(s)=-\gamma x dt+\sqrt{D}dW(s)\quad
\end{equation}
with $D=\gamma=1$.  For later comparison with analytical results,
$x(0)=1$ is used as starting value for the process $x(s)$.

\begin{figure}
  \includegraphics*[width=8cm]{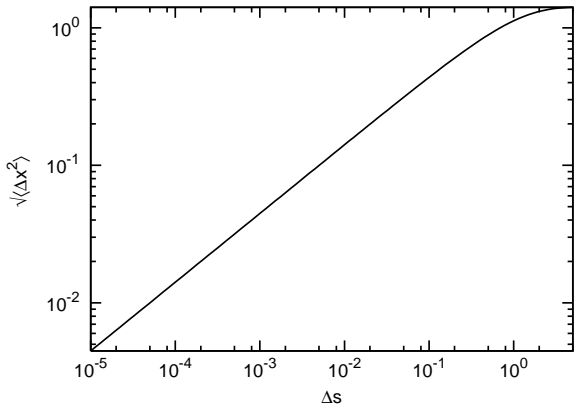}
  \caption{\label{fig:deltas}Determination of the intrinsic increment
    $\Delta s$. From the transition pdf of Ornstein-Uhlenbeck
    processes, that is available in a closed form, the square root of
    the means square deviation of the increment $\Delta x$ as function
    of the increment $\Delta s$ can be derived.  From inspection of
    this graph, $\Delta s=0.0001$ seems to be sufficient for the
    current purpose.}
\end{figure}

For Ornstein-Uhlenbeck processes, joint pdfs for finite time increment
can be calculated in a closed form \cite{Risken}. Then, the statistics
of the increments $\Delta x:=x(s+\Delta s)-x(s)$ can be considered as
a function of the discretization $\Delta s$. The distribution of
$\Delta x$ as a function $\Delta s$ is Gaussian with variance
\begin{equation}
  \left\langle(\Delta x)^2\right\rangle=2\left(1-e^{-\Delta s}\right)\quad.
\end{equation}
In order to estimate an appropriate value for $\Delta s$, the root
mean square deviation has been investigated. From the inspection of
figure \ref{fig:deltas}, $\Delta s=0.0001$ has been selected for
application in the numerical procedure. The maximum deviation of the
definitions (\ref{eqn:invers-def}) from one another is $\sim 10^{-2}$,
which is accurate enough for the current purpose.

\begin{figure*}
  \includegraphics*[width=16cm]{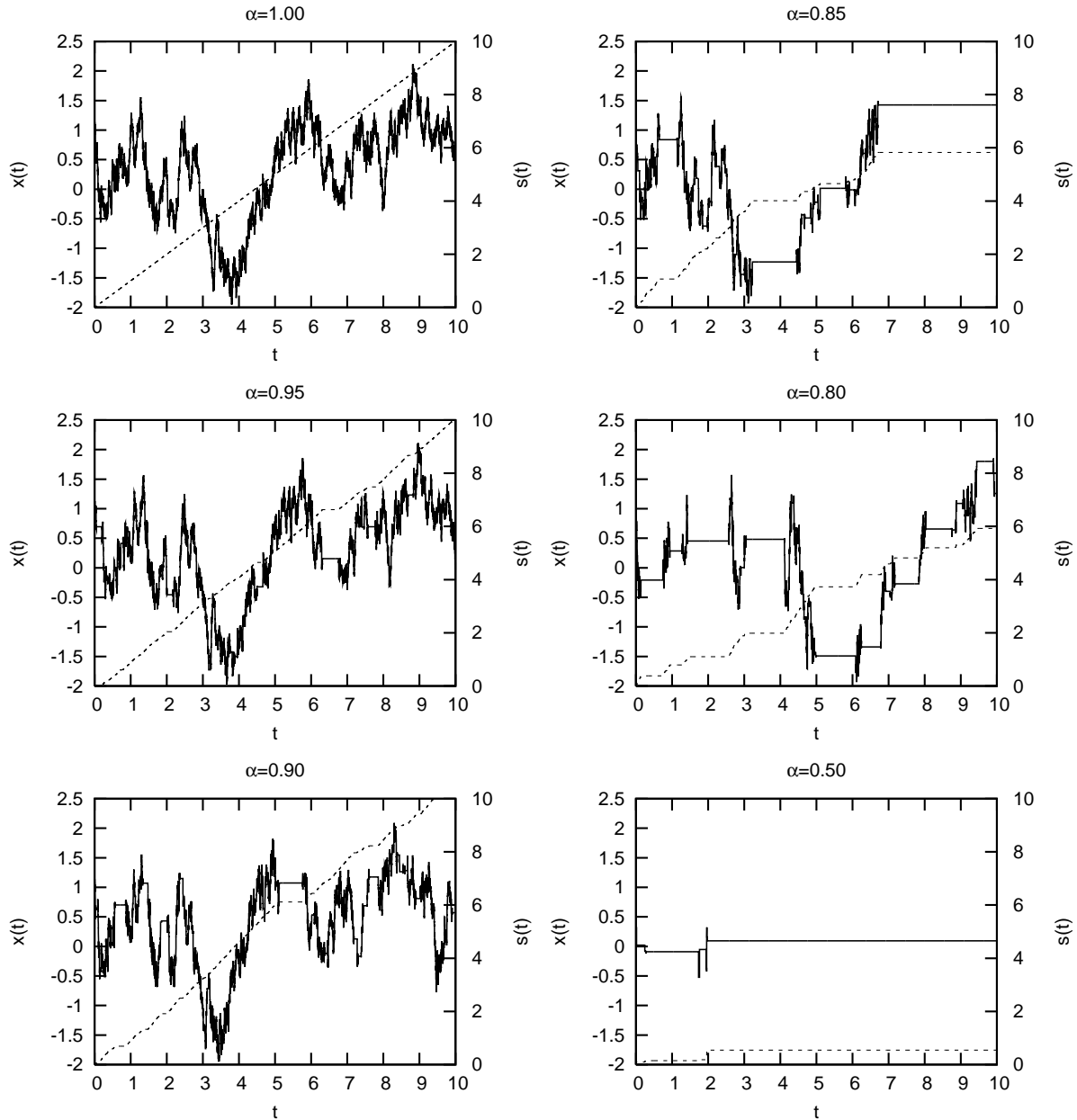}
  \caption{\label{fig:trajekt}Sample trajectories of CTRWs with linear
    repelling force $F(x)=-x$ for different stability indices
    $\alpha$. The solid line corresponds to the process $x(t)$ whereas
    the dashed line indicates the corresponding $s(t)$. $\alpha=1$
    complies with the ordinary Ornstein-Uhlenbeck process. With
    decreasing $\alpha$ the process is dominated by waiting events
    indicated by constant $x$ and $s$.  }
\end{figure*}
\begin{figure}
  \includegraphics*[width=8cm]{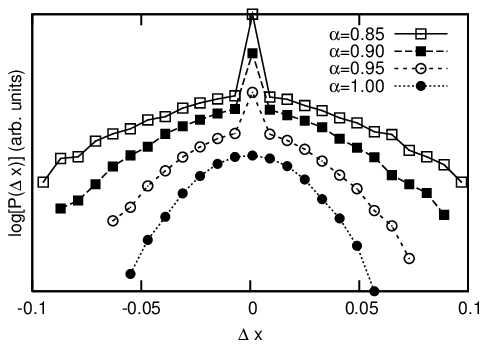}
  \caption{\label{fig:incvert}Increments distributions for $\Delta
    x:=x(t+\tau)-x(t)$ with $\tau=0.001$ for some processes depicted
    in figure \ref{fig:trajekt}. For reasons of clearness the
    individual pdfs are shifted in vertical direction by a constant
    factor. It is evident, that with decreasing stability index
    $\alpha$ a central peaks evolves, that corresponds to persistent
    regions due to waiting events. On the other hand, the
    distributions broaden indicating a higher probability of the
    occurence of extreme increments.  }
\end{figure}

$10000$ data points with time increment $\Delta t=0.001$ have been
generated for several values of $\alpha$.  The trajectories of the
respective processes are exhibited in figure \ref{fig:trajekt}.  From
the sample paths, the influence of the subordinating process $t(s)$ on
the dynamics becomes evident. For $\alpha=1$ an Ornstein-Uhlenbeck
process is recovered. With decreasing $\alpha$ waiting events start to
dominate the process. This also can be seen from the evolution of the
increment pdfs, that is depicted in figure \ref{fig:incvert}.

\begin{figure}
  \includegraphics*[width=8cm]{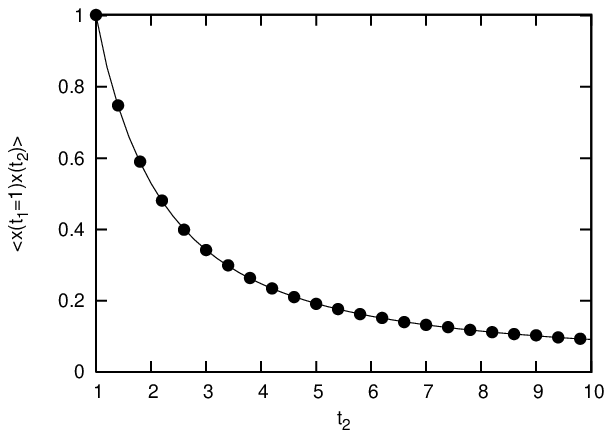}
  \caption{\label{fig:korrelation}Correlation function of the process
    depicted in figure \ref{fig:trajekt} for $\alpha=0.8$. The solid
    line marks the analytical solution (\ref{eqn:corr_eq2}) derived by
    Baule et al.~\cite{Baule07}. The circles indicate the correlation
    obtained from the analysis of an ensemble of $500000$
    trajectories. Since a perfect coincidence is observed, this
    evaluation is proposed as a benchmark for accuracy of the
    numerical implementation.  }
\end{figure}

Recently, the fractional extension of Ornstein-Uhlenbeck processes has
been investigated by Baule et al. \cite{Baule07}, starting from the
fractional Fokker-Planck equation for the time evolution of ensembles
of particles. For the Ornstein-Uhlenbeck process $x(s)$ with initial
value $x(0)=1$, that has been considered in this section, for
$t_2>t_1$ eventually an analytical expression for the correlation
function could be derived,
\begin{eqnarray}
  \label{eqn:corr_eq2}
  &&\langle x(t_2)x(t_1)\rangle=\frac{
    t_1^\alpha}{\Gamma(\alpha+1)}\sum_{n=0}^\infty\frac{(-t_2^\alpha)^n}{\Gamma(\alpha n+1)}\\\nonumber&&\times{_2}F_1\left(\alpha,-\alpha n,\alpha+1;\frac{t_1}{t_2}\right)+E_\alpha(- t_2^\alpha)\quad.
\end{eqnarray}
Here, $\Gamma$ denotes the Gamma function, $E_\alpha$ the
one-parameter Mittag-Leffler function and ${_2}F_1(a,b,c;z)$ the
Gaussian hypergeometric function. For details see the references
provided by Baule et al. \cite{Baule07}.

On the other hand, the correlations can be estimated from ensemble
averages of simulated trajectories $x(t)$. A comparison of the
analytical result by Baule et al. with the correlations estimated from
our simulated trajectories is exhibited in figure
\ref{fig:korrelation}. Perfect coincidence of these two approaches is
observed, that never could be compared before.

\section{\label{sect:conclu}Summary and Conclusions}

\begin{figure}
  \includegraphics*[width=8cm]{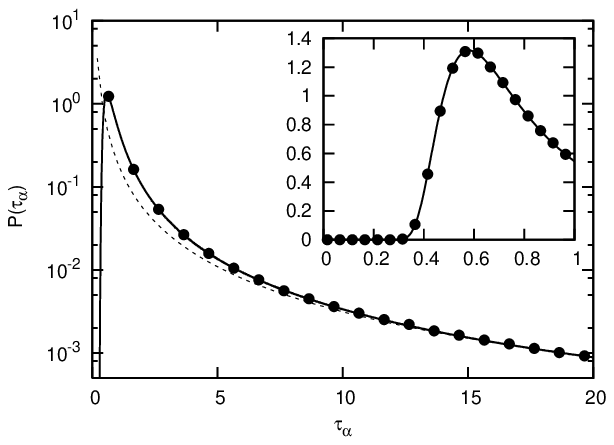}
  \caption{\label{fig:randnum-perform}Performance of the generation of
    skewed $\alpha$-stable random numbers. The solid line exhibits the
    pdf obtained by numerical integration of (\ref{eqn:mylevy}) for
    $\alpha=0.8$. For $x\gg 1$ the pdf shows power law decay with the
    exponent $1+0.8$, that is depicted dashed. The points mark the pdf
    obtained from a sample of $10^{7}$ random numbers, that have been
    generated by means of equation (\ref{eqn:randgen}) for the same
    stability index. The analytical pdf evidently is well reproduced
    by the sample of random numbers.}
\end{figure}

A method for the accurate and efficient simulation of continuous
trajectories of Continuous Time Random Walk (CTRW) processes has been
proposed, that relies on the representation through Langevin equations
proposed by Fogedby. It is based on the simultaneous simulation of two
stochastic processes, one of which is driven by Levy noise.

Within the scope of section \ref{sect:levy-prop}, the construction of
the process $x(t)=x(s(t))$ with the aid of the inverse $s(t)$ of the
Levy process $t(s)$ has been discussed in great detail. The unique
existence of the inverse of $t(s)$ for any $t$ typically has been
assumed in the past \cite{Baule05}. However, the meaning of the
inverse process in fact has to be specified in detail at
discontinuities of $t(s)$ in order to guarantee for unique existence,
see e.g.~equations (\ref{eqn:invers-def}). We would like to emphasize,
that these additional specifications influence neither the
trajectories $x(t)$ nor the ensemble statistics, if the trajectories
$x(s)$ are continuous with probability $1$. In the subsequent
sections, we mainly focussed on this specific case.

Comparison with recent analytical results by Baule et
al.~\cite{Baule07} has been used for validation of the simulation
procedure and showed compliance of the results. Due to the
non-Markovian character of fractional processes, higher order joint
statistics are of great interest \cite{Barkai07}. We propose the use
of probabilistic methods for numerical calculation of these functions.

\section*{Acknowledgements}
The authors kindly acknowledge intensive discussions with Adrian
Baule, Stephan Eule, Michael Wilczek and Eli Barkai.  Adrian Baule
provided the numerical evaluation of equation (\ref{eqn:corr_eq2}),
that has been used in figure \ref{fig:korrelation} for comparison with
our numerical results.  Financial support was granted by the
\emph{Bundesministerium f\"ur Bildung, Forschung und Wissenschaft}
(BMBF) within the project \emph{Windturbulenzen und deren Bedeutung
  f\"ur die Windenergie}.

\appendix

\section{\label{sect:randgen}Generation of random variables with
  skewed $\alpha$-stable pdf according to equation (\ref{eqn:mylevy})}

Skewed Levy-stable random numbers efficiently can be generated by
means of the algorithm proposed in \cite{Weron01, Janicki}. We adapted
this algorithm to our definition of the skewed Levy distributions,
(\ref{eqn:mylevy}).

The random numbers $\tau_{\alpha}(s_i, \Delta s)$, that are required
for numerical integration of the process $t(s)$, then for $0<\alpha\le
1$ efficiently can be generated by means of this algorithm:

\begin{itemize}
\item Generate a random variable $V_i$ uniformly distributed on
  $]-\pi/2,\pi/2[$ and an independent exponential random variable
  $W_i$ with mean $1$. Several optimized random number generators are
  available for this purpose. In case of doubt, $V$ and $W$ can be
  obtained from two independent variables $u^1_i$ and $u^2_i$, that
  are uniformly distributed on $]0,1[$, by means of
  \begin{eqnarray}
    V_i&=&\pi\left(u^1_i-\frac{1}{2}\right)\\
    W_i&=&-\log(u^2_i)\quad.
  \end{eqnarray}
\item Set
  \begin{eqnarray}
    \label{eqn:randgen}
    \tau_{\alpha}(s_i, \Delta s)&=&(\Delta s)^{\frac{1}{\alpha}}\frac{\sin\left[\alpha(V_i+\frac{\pi}{2})\right]}{\left[\cos(V_i)\right]^{(1/\alpha)}}\\\nonumber&&\times\left\{\frac{\cos\left[V_i-¸\alpha(V_i+\frac{\pi}{2})\right]}{W_i}\right\}^{\frac{1-\alpha}{\alpha}}\quad.
  \end{eqnarray} 
\end{itemize}

In case of $\alpha\equiv 1$, $\tau_{1}(s_i, \Delta s)=\Delta s$ is
recovered.  If adequate random number generators are applied for
generation of the variables $V_i$ and $W_i$, the resulting random
numbers $\tau_\alpha$ are uncorrelated. From figure
\ref{fig:randnum-perform} it becomes evident, that the desired skewed
$\alpha$-stable Levy pdf (\ref{eqn:mylevy}) is matched. This algorithm
therefore can be applied for efficient numerical simulation of the
process $t(s)$ according to equation (\ref{eqn:ctrw-fogedby-disc-t}).


\end{document}